\newcount\notenumber
\def\clearnotenumber{\notenumber=0}
\def\note{\advance\notenumber by1 \footnote{$^{\the\notenumber}$}}
\clearnotenumber
\def\e{\rm {e}}
\newcommand{\re}{\ref}
\newcommand{\lab}{\label}
\newcommand{\bu}{\begin{equation}}
\newcommand{\bn}{\begin{eqnarray}}
\newcommand{\eu}{\end{equation}}
\newcommand{\en}{\end{eqnarray}} 
\newcommand{\nn}{\nonumber}

\documentstyle[aps,12pt,preprint,epsfig]{revtex}
\draft
\tightenlines 

\begin{document}

\title{RELATIVISTIC CHARGED SPHERES \\
II: REGULARITY AND STABILITY \thanks{
This work was partially supported by the Gruppo Nazionale per la Fisica
Matematica del CNR and by the Ministero della Ricerca Scientifica e
Tecnologica (MURST) of Italy. \dag: defelice@pd.infn.it, \ddag: liusiming@263.net, \S: yqyu@hotmail.com.}  }
\author{Fernando de Felice \dag, Liu Siming \ddag  and Yu Yunqiang \S \\ 
\dag {\small Department of Physics {\it G. Galilei}, University of Padova} \\
{\small via Marzolo 8, I-35131 Padova Italy and INFN, Sezione di Padova}\\
\ddag \S {\small Department of Physics, Peking University, Beijing 100871, China } }

\date{\today}
\maketitle
\begin{abstract}
We present new results concerning the existence of static, 
electrically charged, perfect fluid spheres that have a regular interior 
and are arbitrarily close 
to a maximally charged  black-hole state. These configurations are described 
by exact solutions of Einstein's field equations. A family of these solutions 
had already be found (de Felice et al., 1995) but
here we generalize that result to cases with different charge distribution 
within the spheres and show, in an appropriate parameter space, that the 
set of such physically reasonable solutions has a non zero measure. 
We also perform a perturbation analysis and identify the solutions 
which are stable against adiabatic radial perturbations. We then suggest 
that the stable configurations can be considered as classic models of 
charged particles. Finally our results are used  to show that a conjecture of 
Kristiansson et al. (1998) is incorrect. 
\end{abstract}

\bigskip   


\section{Introduction}

It is well known that static, spherically symmetric,
uncharged perfect fluids cannot be held in equilibrium below a certain 
radius
without developing singularities inside.
This radius is $9M/4$ for an incompressible fluid sphere of total mass 
energy $M$;
it can be even larger than that for more realistic equations of state 
(Buchdahl, 1959).
The possibility of  holding  a non-singular object in stable equilibrium but 
compact enough to be close (in fact arbitrarily close) to a black-hole 
state, 
is of great interest not 
only in order to judge the state of matter in this quasi-critical condition, 
that is being about to turn into a black-hole,
but also to yield  a classic model of charged massive particles 
which might have 
astrophysical and cosmological implications.
Although one can reach this goal with non-perfect fluids, a perfect fluid 
solution of the type 
mentioned was recently found (de Felice et al., 1995) but with the 
presence of an electric charge. Electric charges inhibit the growth of 
 space-time 
curvature and therefore they are an efficient means of avoiding 
singularities inside 
matter and enhance stability even for quasi-critical configurations.

In this paper 
we extend the above mentioned  analysis (de Felice et al., 1995) to a 
more general case
and, in Section 2, show
the existence of regular solutions very close to a black-hole state as 
points in a parameter space (figures \re{Fig1}-\re{Fig3}).
In Section 3 we study the stability of these solutions against 
adiabatic radial 
perturbations and show the domain of stable solutions in the same 
parameter space (figures \re{Fig1}, \re{Fig2}). Having solutions which describe very compact 
sources, we are able to verify the correctness of a conjecture  
(Kristiansson et al., 1998) according to which the limit of regular 
embedding in the Euclidean space of conformally reduced space-like sections 
of the 
Reissner-Nordstr\"om metric, coincides with the limit of regularity of 
the internal solutions. In section 4 we show that this is not true.

Throughout the paper we use 
geometrized units ($c=1=G$) and metric signature $+2$.

\section{Existence of regular quasi-critical spheres}

In Schwarszchild coordinates the line element of a spherically symmetric 
space-time reads:
$$
ds^2=-\e^{\eta}dt^2+\e^{\lambda}dr^2+r^2d\theta^2+r^2\sin^2\theta d\phi^2
$$
where $\eta=\eta(r,\,t)$ and $\lambda=\lambda(r,\,t)$. 
The energy-momentum tensor of a perfect, electrically 
charged fluid takes the form (de Felice et al.,1990):
$$
T^{\mu\nu}=(\rho+p)u^\mu u^\nu +pg^{\mu\nu}+
{1\over 4\pi}(F^{\mu\alpha}F^\nu{}_\alpha-
{1\over 4}g^{\mu\nu}F_{\alpha\beta}F^{\alpha\beta})
$$
where $F^{\alpha\beta}$ satisfies Maxwell's equations:
$$
F_{[\alpha\beta,\gamma]}=0,\qquad F^{\mu\nu}{}_{;\nu}=4\pi j^\mu.
$$
Here $j^\mu$ is the current four-vector, $\rho$ and $p$ are the 
energy density and the isotropic pressure 
of matter measured in its rest frame respectively; $\rho$ and $p$ 
are expressed in (geometrized) units of $length^{-2}$.

In the spherically symmetric case the only non vanishing component 
of Maxwell's tensor is 
(Bekenstein, 1971): 
$$
 F^{tr}=\e^{-{\lambda+\eta\over 2}}{Q(r,t)\over r^2}
$$
where:
\bu
Q(r,t)=\int_0^r\e^{\lambda+\eta\over 2}4\pi r^2 j^t dr
\lab{y5}
\eu
is the total 
electric charge within a sphere of radius $r$ at time $t$; it 
then follows that: 
$$F^{\alpha\beta}F_{\alpha\beta}=-{2Q^2\over r^4}$$
where charge is expressed in units of $length$.
If we require that the configuration be static, then $u^\mu=\e^{-\eta/2}
\delta^\mu_t$, hence Einstein's equations which one needs to solve 
for the interior metric, take the form:
\bn
\e^{-\lambda}\left({\lambda'\over r}-{1\over r^2}\right)+{1\over r^2}
&=&{Q^2\over r^4}+8\pi\rho,   \lab{y1}   \\
-\e^{-\lambda}\left({\eta'\over r}+{1\over r^2}\right)+{1\over r^2}
&=&{Q^2\over r^4}-8\pi p 
\lab{y6}
\en
where $(')$ means derivative with respect to $r$.  

Let the sphere be incompressible with a constant matter energy density $\rho_m$ and 
the charge distribution within the sphere be given by:
\bu
Q(r)=Q_0\left({r\over R}\right)^n
\lab{y7}
\eu
where $n$ is a constant parameter, $R$ is the coordinate radius of the 
boundary and $Q_0$ is the total charge. Integrating equation (\re{y1}) with respect to coordinate radius $r$, 
one obtains
\bu
\e^{-\lambda}=1-{8\pi\over 3}\rho_m r^2-{{Q_0}^2r^{2n-2}\over (2n-1)R^{2n}}. \lab{y2}
\eu
Combining this result with (\re{y5}) and (\re{y7}), we see that the 
charge density is given by:
$$
\e^{\eta\over 2}j^t={Q_0nr^{n-3}\over4\pi R^n\e^{\lambda\over 2}}
={Q_0nr^{n-3}\sqrt{1-{8\pi\over 3}\rho_m r^2-{{Q_0}^2r^{2n-2}\over (2n-1)R^{2n}}}\over4\pi R^n},
$$
then we have $n\ge 3$ which prevents the divergence of charge density 
at the center. 

Introducing scale parameters such as:
\bn
p_0={1\over 3}\rho_m, \qquad r_0=\left({3\over 4\pi\rho_m}\right)^{1/2} 
\lab{y9}
\en
and setting:
\bu
Y\equiv {p\over p_0},\qquad \bar\xi\equiv {r\over r_0}
\lab{y10}
\eu
the Oppenheimer-Wolkoff equation and equation (\re{y2}) for such a 
sphere read respectively (de Felice et al., 1995):
\bn
{dY\over d\bar\xi}&=&n\alpha\bar\xi^{2n-5}-(Y+3){(Y+1)\bar\xi-
[(n-1)/(2n-1)]\alpha\bar\xi^{2n-3}\over 1-2\bar\xi^2-
[\alpha/(2n-1)]\bar\xi^{2(n-1)}}, \lab{y11} \\
e^{-\lambda}&=&1-2\bar\xi^2-[\alpha/(2n-1)]\bar\xi^{2(n-1)} \lab{y3}
\en
where
$$\alpha=\left({Q_0r_0^{n-1}\over R^n}\right)^2. $$ 
An obvious condition  is $\alpha>0$. Now we see that the set of equations (\re{y6}), (\re{y11}) and (\re{y3}) is complete.

\begin{figure}[\ht]
\vspace{-0.2cm}
\centerline{\epsfig{figure=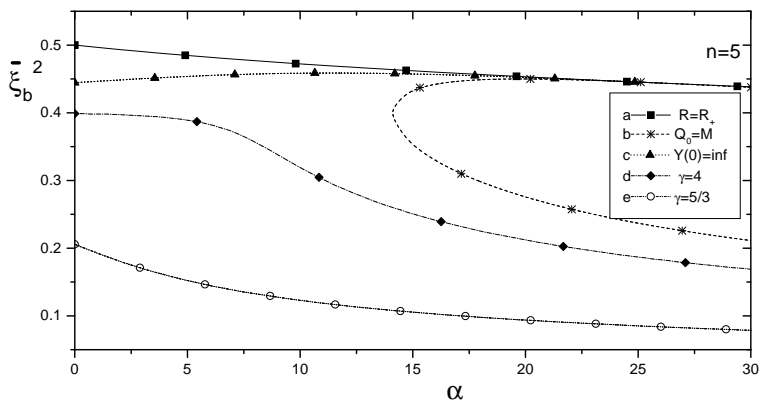, height=7cm}}
\vspace{-0.5cm}
\caption{Configurations in the parameter space for n=5}
\label{Fig1}

\end{figure}
The boundary of the 
physical configuration is identified by the conditions:
\bu
Y(\bar\xi_b)=0, \quad e^{\nu}(\bar\xi_b)=e^{-\lambda}(\bar\xi_b)=1-2{\bar\xi_b}^2-{\alpha\over 2n-1}{\bar\xi_b}^{2(n-1)}.
\lab{y12}
\eu
where $\bar\xi_b=R/r_0$. Then, for a given set of $(\alpha,\ \bar\xi_b,\ n)$, 
one can solve equations (\re{y6}), (\re{y11}) and (\re{y3}) and obtain the structure 
of the corresponding configuration (Note: When one integrates equation (\re{y11}) 
numerically from its boundary inward, pressure $Y$ may diverge before $\bar\xi$ 
reaches zero for some values of $(\alpha,\ \bar\xi_b,\ n)$. Since there is a
singularity in such a configuration, it is physically meaningless. But we still use 
the parameter $(\alpha,\ \bar\xi_b,\ n)$ to represent such solutions). 
But not all of the solutions are acceptable, two basic requirements of a regular static configuration are:

1: That there be no singularity inside the sphere,

2: That the radius of the boundary be larger than the external horizon size of the black hole with the same mass and charge. 

\begin{figure}[\ht]
\vspace{-0.20cm}
\centerline{\epsfig{figure=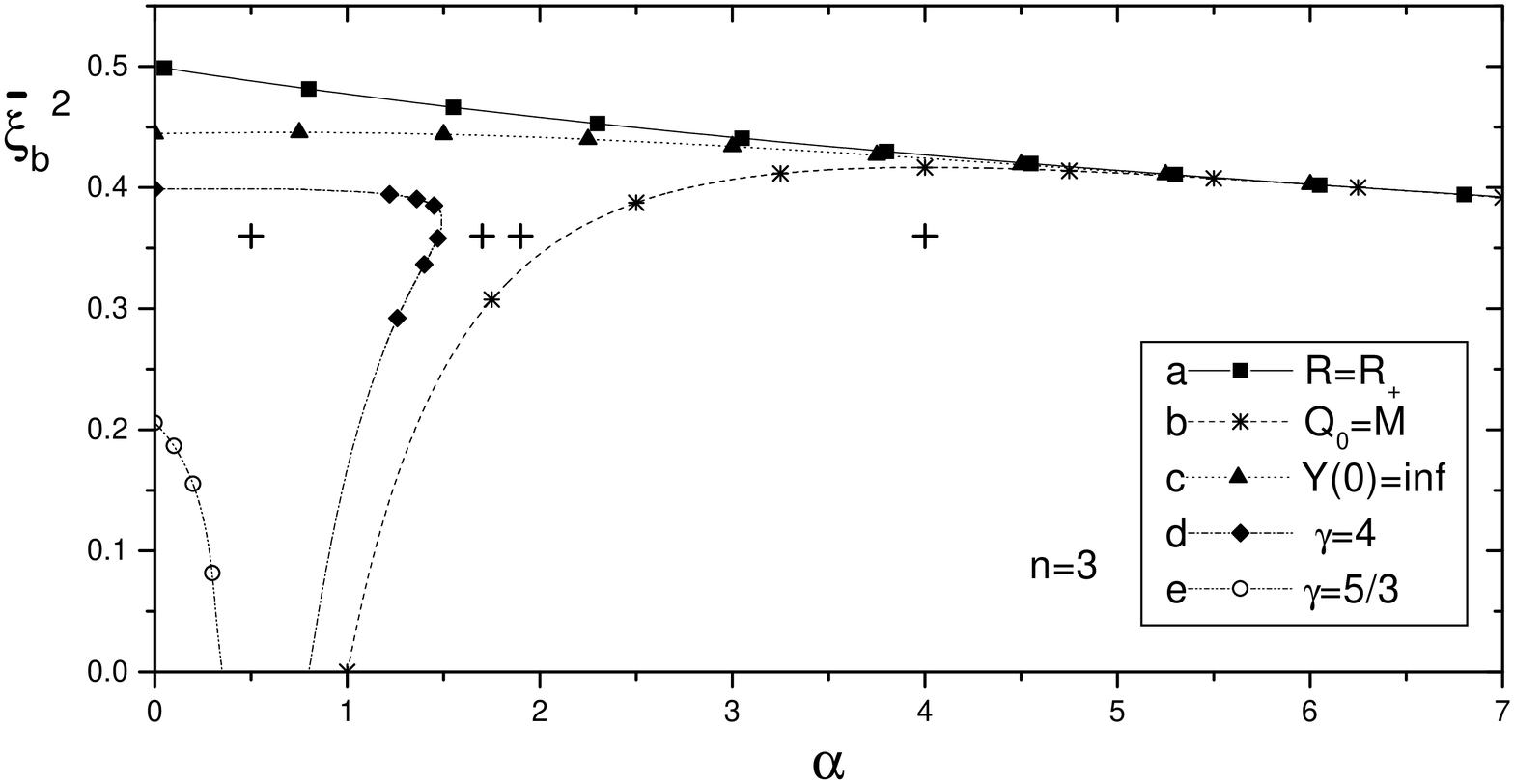, height=7cm}}
\vspace{-0.50cm}
\caption{Configurations in the parameter space for n=3}
\label{Fig2}
\vspace{0.1cm}
\centerline{\epsfig{figure=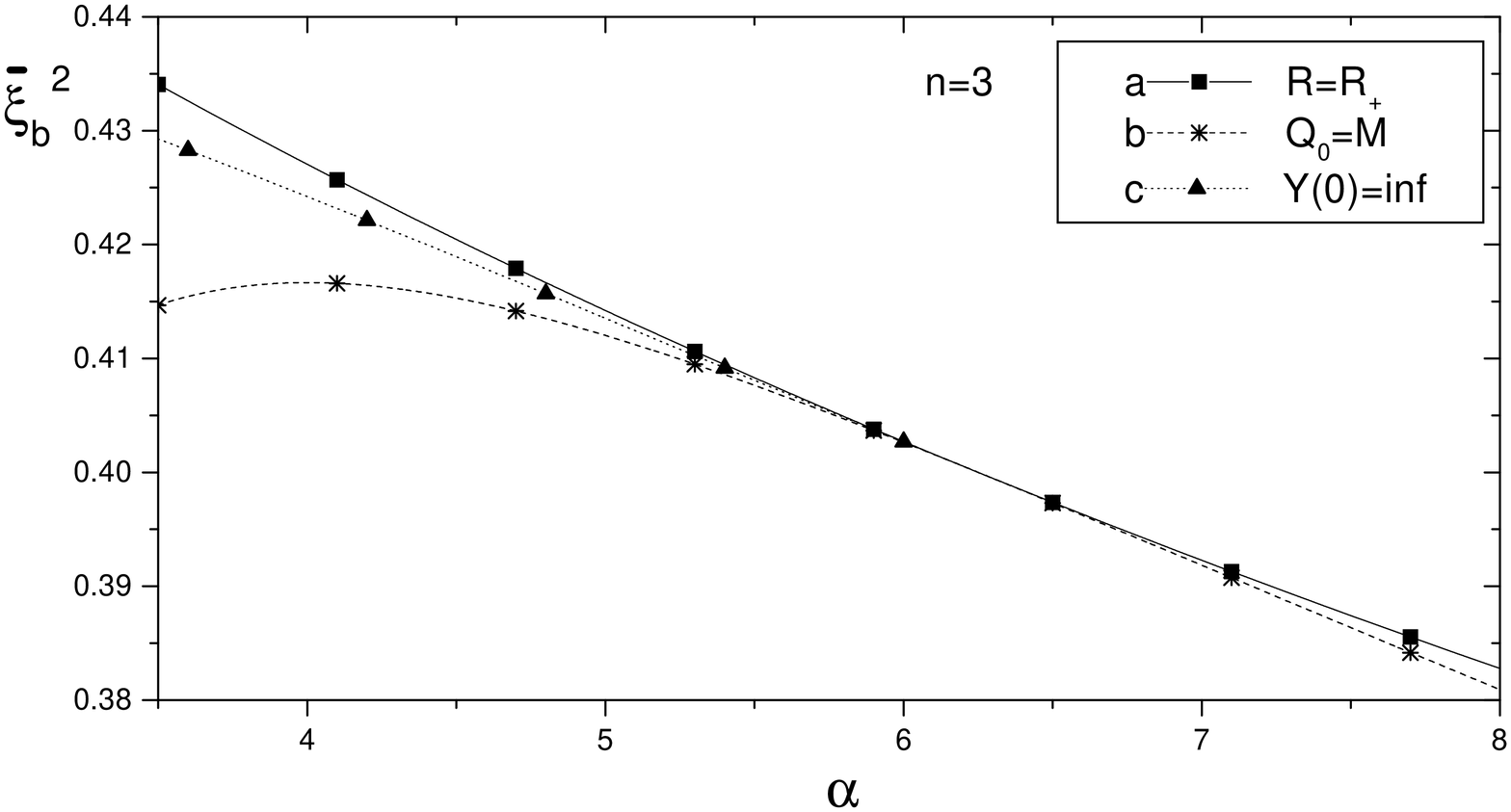, height=7cm}}
\vspace{-0.50cm}
\caption{Enlargement of figure \re{Fig2} around the meeting point}
\label{Fig3}
\end{figure}

For a given $n$, each point in the $(\bar\xi_b^2,\,\alpha)$-plane represents a solution. In figures \re{Fig1}-\re{Fig3} we plot the functions which allow 
one to single out the points which correspond to physical configurations. 

Let $M$ be the total mass energy of the solution, then the 
relations between $\bar\xi_b$ and the other parameters are:
\bn
{R\over M}&=&{1\over \bar\xi_b^2\left[1+\bar\xi_b^{2(n-2)}
{n\alpha
\over 2n-1}\right]}, \lab{y13} \\
{Q_0\over M}&=&{\sqrt\alpha\over \bar\xi_b^{3-n}\left[1+\bar\xi_b^{2(n-2)}
{n\alpha
\over 2n-1}\right]}. 
\lab{y14}
\en
One can show that $R/M$ changes monotonically along the curve $Q_0/M=constant$. 
Let $Q_0=M$, then (\re{y14}) gives the curve $b$ where $Q_0=M.$ Inside (figure \re{Fig1}, for $n>3$) or below (figure \re{Fig2}, for $n=3$) this curve, 
$Q_0>M$.  

The horizon sizes for the charged black hole are defined as the solutions of 
equation $e^{\lambda}(R)=0$ and the results are 
$R_{\pm}=M\pm\sqrt{M^2-{Q_0}^2}$ for a 
given $M$ and $Q_0$. Let $e^{\lambda}(\bar\xi_b)=0$, (\re{y3}) gives  curve $a$ 
where $R=R_{+}.$ Above this curve, we have $R<R_+$, so the solutions are unacceptable.
We notice this curve is tangential to the curve $Q_0=M$ at the point 
$\bar\xi_b^2=(n-1)/(2n-1),\,\alpha=[(n-1)/(2n-1)]^{1-n}$. On the left hand side of 
this point, curve $a$ represents configurations with their boundary at the external 
horizon. Consequently  physical configurations must lie below this curve.

For given $\alpha$ and ${\bar\xi_b}^2$, one can solve equation 
(\re{y11}) numerically to get $Y(\bar\xi)$. We find that the central 
pressure $Y(0)$ will diverge as ${\bar\xi_b}^2$ increases to some value for 
fixed $\alpha<[(n-1)/(2n-1)]^{1-n}$, then we get the curve $c$ $Y(0)=inf$ 
which approaches the tangential point of curves $a$ and $b$. 
Above this curve, there are pressure singularities inside the configurations. 
So the regular static configurations must lie below this curve. (Curves $d$ 
and $e$ are for later use).

We see from the figures that, for these special models,  no regular solution exists which describes a compact 
charged sphere having $Q_0<M$ and radius $R$ arbitrarily close to 
$R_+$ since 
any such solution would lie above curve $c$ where  a
pressure singularity emerges inside the sphere. 
A more general result is assured by the following:

\medskip
{\it Theorem}: if the total electric charge of a static perfect fluid ball  
is smaller than 
its total mass, then there is no regular static configuration having a radius  
arbitrarily close to the size of the external horizon (Yu et al., 1999).
\medskip

As stated in (de Felice et al., 1995), the curves in figures \re{Fig1}-\re{Fig3} show 
that a finite
region in the parameter space exists, which is the region between curves $b$ and $c$, 
where regular  
solutions describing static charged balls of matter may be found
with a compactness 
arbitrarily close the critical one for transition to a black-hole state, 
namely with $R\to R_+$ and $Q_0\to M$.  

\begin{figure}[\ht]
\vspace{-0.4cm}
\centerline{\epsfig{figure=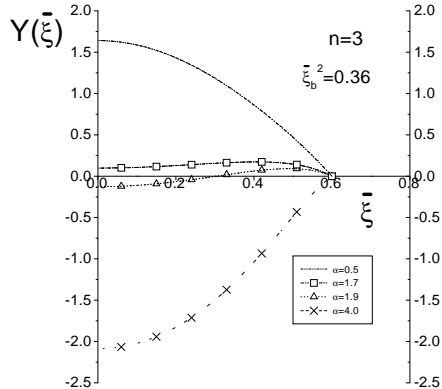, height=7cm}}
\vspace{-0.5cm}
\caption{The pressure of the configurations for n=3}
\label{Fig4}
\end{figure}

It is of interest to study the behavior of the pressure 
inside some of these configurations as a function of parameter $\alpha$.
For a fixed $n$, this parameter gives a measure of the total charge 
within the body, so we show in figure \re{Fig4} plots of
the internal pressure as function of radius $\bar\xi$, for $n=3$, $\bar\xi_b^2=0.36$ and 
increasing values of $\alpha$ (the corresponding configurations are indicated by `+' 
in figure \re{Fig2}). 
It is interesting  to see how charge contributes to a decrease of the 
central pressure, a property which justifies
the very existence of regular solutions describing  compact balls of matter.

Regular solutions also exist inside (for $n>3$) or 
below (for $n=3$) curve $b$ in figures 1 and 2, where $Q_0>M$. However, as  appears 
from figure 4, the central pressure is negative in these solutions.

In what follows we shall discuss the stability
of these solutions against adiabatic radial perturbations.

\section{Stability}

We will consider small radial perturbations for the 
charged, perfect fluid balls, then any fluid element which was at $r$ in the 
unperturbed configuration is displaced
to $r+\xi(r,t)$ in the perturbed one where $\xi\ll r$. In what follows we 
shall use 
subscript "$i$" to denote the unperturbed configuration, then we have
(Misner et al., 1973):
\bn
\eta(r,t)&=&\eta_i(r)+\delta\eta(r,t) \nn \\
\lambda(r,t)&=&\lambda_i(r)+\delta\lambda(r,t) \nn \\
p(r,t)&=&p_i(r)+\delta p(r,t) \nn \\
\rho(r,t)&=&\rho_i(r)+\delta\rho(r,t) \nn \\
\bar n(r,t)&=&\bar n_i(r)+\delta \bar n(r,t) 
\lab{y15}
\en
where $\bar n$ is the baryon number density; the variations are small
and such that $\delta a/a_i\sim \xi/r$, $a$ being any of the above 
parameters.
Furthermore we assume that the 
charge distribution in the vibrating configuration remains unchanged, 
namely 
 $Q(r+\xi(r,t))=Q_i(r)$,
meaning that there are no electric currents for the comoving observer.
To first order in the variations, the Lagrangian perturbations read:
\bn
\Delta p(r,t)&=&p(r+\xi,t)-p_i(r)\cr
                          &=&\delta p(r,t)+{p_i}'\xi\cr
           \Delta\rho(r,t)&=&\delta\rho(r,t)+{\rho_i}'\xi \cr
         \Delta\bar n(r,t)&=&\delta\bar n(r,t)+{\bar n_i}'\xi  \cr
\Delta Q(r,t)&=& 0. 
\lab{y16}
\en

From baryon conservation we have:
\bu
{d\Delta \bar n\over d\tau}=-\bar nu^\mu{}_{;\mu}.
\lab{y17}
\eu
Since $u^r/u^t=\dot\xi$, the dot meaning differentiation with respect to the 
coordinate time $t$ and $u^\mu u_\mu=-1$,  in the 
perturbations we have to first order:
\bn
u^t&=&\e^{-\eta_i/2}\left(1-{\delta\eta\over 2}\right)\cr
u^r&=&\e^{-\eta_i/2}\dot\xi. 
\lab{y18}
\en
Using the relation
$$ u^\mu{}_{;\mu}=(-g)^{-1/2}\left[(-g)^{1/2}u^\mu\right]_{,\mu}$$
and integrating equation (\re{y17}) with respect to coordinate time (first-order analysis), we deduce:
\bu
\Delta\bar n=-\bar n_i\left[r^{-2}
\e^{-\lambda_i/2}\left(r^2\e^{\lambda_i/2}\xi\right)'+
{\delta\lambda\over 2}\right]. 
\lab{y19}
\eu

For adiabatic variations we have:
$$ {\Delta p\over\Delta\bar n}={p_i\over \bar n_i}\gamma$$
where $\gamma$ is the adiabatic index, then from (\re{y19}) and (\re{y16})
we get:
\bu
\delta p=-\gamma p_i
\left[r^{-2}\e^{-\lambda_i/2}\left(r^2\e^{\lambda_i/2}\xi\right)'+
{\delta\lambda\over 2}\right]-\xi{p_i}' .
\lab{y20}
\eu

From Einstein's equations, we obtain:
\bn
-e^{-\lambda_i}{r{\eta_i}'+1\over r^2}\delta\lambda+
\e^{-\lambda_i}{(\delta\eta)'\over r}&=&
{2Q_i{Q_i}'\over r^4}\xi+8\pi\delta p \lab{y25} \\
\e^{-\lambda_i}{\dot{(\delta\lambda)}\over r}\ =\ -8\pi(\rho_i+p_i)\dot\xi
&=&-{\e^{-\lambda_i}\over r}({\lambda_i}'+{\eta_i}')\dot\xi 
\lab{y27}
\en
thus, integrating (\re{y27}) with respect to coordinate time and choosing the integration 
constant so that $\delta \lambda=0$ when $\xi=0$, we get:
\bu
\delta\lambda=-\e^{\lambda_i}8\pi r(\rho_i+p_i)\xi=-({\lambda_i}'+
{\eta_i}')\xi.
\lab{y28}
\eu
Using (\re{y28}), (\re{y20}) and (\re{y6}) to eliminate respectively $\delta\lambda$, 
$\delta p$ and $\eta_i'$ in (\re{y25}), we have:
\bn
(\delta\eta)' &=&\e^{\lambda_i}r\left({2Q_i{Q_i}'\over r^4}\xi+
8\pi\delta p\right)+{r{\eta_i}'+1\over r}\delta\lambda \cr
&=&-8\pi\gamma p_i{\e^{\lambda_i+\eta_i/2}\over 2}\left(r^2\e^{-\eta_i/2}\xi
\right)'+
8\pi[{p_i}' r-(p_i+\rho_i)]\e^{\lambda_i}\xi-{2Q_i{Q_i}'\over r^3}
\xi\e^{\lambda_i}.
\lab{y29}
\en

If we project the identity:
\bu
T^\nu{}_{\mu;\nu}=0
\lab{y21}
\eu
parallelly to the matter flow, namely:
$$ u^\mu T^\nu{}_{\mu;\nu}=0,$$
we obtain, from the chosen forms of $u^\mu$ and $T^{\mu\nu}$:
\bu
-{d\rho\over d\tau}+{(\rho+p)\over\bar n}{d\bar n\over d\tau}-
{Q\over 4\pi r^4}{dQ\over d\tau}=0.
\lab{y22}
\eu
Integrating (\re{y22}) with respect to time and using (\re{y16}) and  (\re{y19})  
to remove $\Delta Q$, $\Delta\rho$ and 
$\Delta \bar n$ we deduce after some algebra:
\bu
\delta\rho=-(\rho_i+p_i)\left[r^{-2}\e^{-\lambda_i/2}
\left(r^2\e^{\lambda_i/2}\xi\right)'+
{\delta\lambda\over 2}\right]-\xi{\rho_i}'.
\lab{y23}
\eu
If we now project (\re{y21}) transversely to the matter flow, namely:
$$ h^\mu{}_\nu T^\sigma{}_{\mu;\sigma}=0,$$
where $h^\mu{}_\nu=\delta^\mu_\nu+u^\mu u_\nu$, we obtain:
\bu
(\rho+p)u^\mu u_{\nu;\mu}=-p_{,\nu}-{dp\over d\tau} 
u_\nu-(T^\mu{}_{\nu;\,\mu})^{em}-u_\nu{Q\over 4\pi r^4}
{dQ\over d\tau}.
\lab{y30}
\eu
The $r$-component of (\re{y30}) yields:
\bu
(\rho_i+p_i)\e^{\lambda_i-\eta_i}\ddot\xi=-(\delta p)'-
(\delta\rho+\delta p)
{{\eta_i}'\over 2}-
(\rho_i+p_i){(\delta\eta)'\over 2}-{Q_i{Q_i}'\over 4\pi r^4}\xi'-
{Q_i^{\prime 2}\over 4\pi r^4}\xi ,
\lab{y31}
\eu
while the $t$-component is trivial and gives:
$${p_i}'={Q_i{Q_i}'\over 4\pi r^4}-(\rho_i+p_i){{\eta_i}'\over 2}.$$

By using the initial value equations (\re{y20}), (\re{y23}) and (\re{y29}) 
to reexpress $\delta p$, 
  $\delta\rho$ and $(\delta\eta)'$ in terms of $\xi$ 
in equation (\re{y31}), we obtain:
\bn
(\rho_i+p_i)\e^{{3\over 2}\lambda_i}\ddot\xi&=& 
\left\{\left[\gamma p_ir^{-2}\e^{\eta_i/2}\zeta'+\xi{p_i}'\right]'+\left[
(\rho_i+p_i(1+\gamma))r^{-2}\e^{\eta_i/2}\zeta'+\xi({\rho_i}'+{p_i}')
\right]{{\eta_i}'\over 2}\right.
\cr
& &+(\rho_i+p_i)\left[4\pi\gamma p_ir^{-1}\e^{\lambda_i+\eta_i/2}\zeta'+{
Q_i{Q_i}'\over r^3}\xi\e^{\lambda_i}+4\pi(\rho_i+p_i)\e^{\lambda_i}\xi-
4\pi{p_i}'r\e^{\lambda_i}\xi\right]
\cr
& &\left.-{Q_i{Q_i}'\over 4\pi r^4}\xi'-{{Q_i}^{\prime 2}\over 4\pi r^4}\xi
\right\}
\e^{\eta_i+\lambda_i/2}
\cr
&=& \left[\gamma p_i r^{-2}\e^{\lambda_i/2+3\eta_i/2}\zeta'\right]'+
\e^{\lambda_i/2+3\eta_i/2}\zeta \left[{(p_i^\prime-
(Q_iQ_i^\prime)/(4\pi r^4))^2
\over r^2(\rho_i+p_i)}\right.\cr
& & \left.+{4\over r^3}\left({Q_iQ_i^\prime\over 
4\pi r^4}-p_i^\prime\right)
-8\pi p_i(\rho_i+p_i){\e^{\lambda_i}\over r^2}-{\e^{\lambda _i}\over 
r^6}Q_i^2(\rho_i+p_i)\right.\cr
& &\left.+{Q_iQ_i^{\prime\prime}\over 4\pi r^6}-
{Q_iQ_i^\prime\over\pi r^7}\right]
\lab{y32}
\en
where $\zeta=r^2\e^{-\eta_i/2}\xi$ is a 
{\it renormalized displacement function}. From the above, the 
physically acceptable solutions of the dynamic equation
must satisfy the following boundary conditions:
\bn
{\xi\over r}\qquad {\rm{finite\, \ or\,\  zero\ \, as}}
\qquad r&\to 0&  \lab{y33} \\
\Delta p=-\gamma p_i r^{-2}\e^{\eta_i/2}
\left(r^2\e^{-\eta_i/2}\xi\right)'\to 0\qquad {\rm{as}}
\qquad r&\to R &
\lab{y34}
\en
where $R$ is the radius of the ball. 

Assume that $\zeta(r,t)$ has a sinusoidal time dependence:
\bu
\zeta(r,t)=\zeta(r)\e^{-i\omega t}
\lab{y35}
\eu
hence the dynamic equation and the boundary conditions reduce to an 
eigenvalue 
problem for the angular frequency $\omega$ and amplitude $\zeta(r)$:
\bu
(F\zeta')'+(H+\omega^2W)\zeta=0
\lab{y36}
\eu
so that $\zeta/r^3$ is finite or zero as $r\to 0$ and 
$\gamma p_i r^{-2}\e^{\eta_i/2}\zeta^\prime(r)\to 0$ as $r\to R$, where:
\bn
F&=&\gamma p_i r^{-2}\e^{\lambda_i/2+3\eta_i/2} \lab{y37} \\
H&=&\e^{\lambda_i/2+3\eta_i/2}\left[
{(p_i^\prime-(Q_iQ_i^\prime)/(4\pi r^4))^2\over r^2(\rho_i+p_i)}
+{4\over r^3}\left({Q_iQ_i^\prime\over 4\pi r^4}-
p_i^\prime\right)-8\pi p_i(\rho_i+p_i){\e^{\lambda_i}\over r^2}\right. \cr
& &\left.-{\e^{\lambda _i}\over r^6}Q_i^2(\rho_i+p_i)+{Q_iQ_i^{\prime\prime}
\over 4\pi r^6}-{Q_iQ_i^\prime\over\pi r^7}\right] \lab{y38} \\
W &=& (\rho_i+p_i)r^{-2}\e^{3\lambda_i/2+\eta_i/2} 
\lab{y39}
\en
By using the variational principle (Mathews J. and Walker R.L., 1970), one gets:
\bu
\omega^2={\rm {extreme\,\  of}}\qquad{\int_0^R(F\zeta^{'2}-
H\zeta^2)dr\over\int_0^RW\zeta^2 dr}
\lab{y40}
\eu
so, the configuration is stable against adiabatic radial perturbations only if:  
\bu
\int_0^R (F\zeta^{\prime 2}-H\zeta^2) dr>0
\lab{y41}
\eu
for any $\zeta$ which satisfies the 
boundary conditions (\re{y33}), (\re{y34}).

Using the simplest $trial\ \ function$ $\zeta\propto r^3$ 
(Chandrasekhar, 1964) 
and assuming that the adiabatic index $\gamma$ was constant throughout the spheres,  
we calculated the stability of such configurations. 
The domain of stable solutions in the parameter space is shown in figures \re{Fig1} and \re{Fig2}:

curve $d$  for $\gamma$=4, the configurations are unstable above this curve,

curve $e$ for $\gamma$=5/3, the configurations are unstable above this curve,

\noindent and the stability curves approach curve $c$ as the adiabatic index 
approaches infinity. Evidently this limit 
corresponds to perfect rigidity.

With the inclusion of these curves, figures \re{Fig1} and \re{Fig2} allow us to identify a region
on the parameter space which corresponds to solutions which are 
physically significant, being regular and stable against adiabatic radial perturbations. 
Since these very compact configurations are electrically charged, they 
are probably  not describing astrophysical objects but, 
perhaps, classic  models of massive elementary particles. 
In this case it would be desirable to extend our analysis to a more detailed 
moldelling of the interior, since the adiabatic index depends critically on 
the equation of state (Akmal et al., 1998).

\section{Regular solutions and regular embeddings}

Given a Riemannian manifold $M$ with Lorentzian metric $g$, 
it may be heuristically helpful to find embedding diagrams of any subset 
of $M$ into  Euclidean space. 
More specifically, one has to find a surface in a three-Euclidean space which
is isometric to a given two-surface of $M$ ( Bini  et al., 1999). 
This problem does not always have a solution.
It has been argued (Kristiansson et al., 1998) that the limiting 
condition for regular embedding of 
space-like sections of the
vacuum
Schwarzschild and Reissner-Nordstr\"om solutions, 
conformally reduced by the factor $\Omega=g_{00}$, does coincide with the 
limiting condition for their regular,
static, spherically symmetric internal solutions.
We shall show here that this is not true.

Consider the Reissner-Nordstr\"om space-time solution:
\bu
ds^2=-\left(1-{2m\over r}+{Q_0^2\over r^2}\right)dt^2+
\left(1-{2m\over r}+{Q_0^2\over r^2}\right)^{-1}dr^2+
r^2(d\theta^2+\sin^2\theta d\phi^2)
\lab{y42}
\eu
then consider the conformal $t=$constant space-like subspace:
\bu
d\Sigma^2=\left(1-{2m\over r}+{Q_0^2\over r^2}\right)^{-2}dr^2+
\left(1-{2m\over r}+{Q_0^2\over r^2}\right)^{-1}r^2(d\theta^2+\sin^2\theta d\phi^2)
\lab{y43}
\eu
The two-dimensional surface $\theta=\pi/2$, namely:
\bu
d\sigma^2=\left(1-{2m\over r}+{Q_0^2\over r^2}\right)^{-2}dr^2+
\left(1-{2m\over r}+{Q_0^2\over r^2}\right)^{-1}r^2d\phi^2
\lab{y44}
\eu
is matched to a surface $z=z(R)$ in the three-dimensional Euclidean space:
\bn
d\sigma_e^2&=&dR^2+R^2d\phi^2+dz^2  \cr
&=&\left[1+\left({dz\over dR}\right)^2\right]dR^2+R^2 d\phi^2 
\lab{y45}
\en
where $R,\, \phi, \, z$ are cylindrical coordinates.
Comparing (\re{y45}) with (\re{y44}) we obtain the condition for regular embedding as:
\bu
\left({dz\over dR}\right)^2=\left(1-{2m\over r}+{Q_0^2\over r^2}\right)
\left(1-{3m\over r}+{2Q_0^2\over r^2}\right)^{-2}-1\geq 0
\lab{y46}
\eu
From this we deduce, for each $M$ and $Q_0$, the condition 
for regular embedding as $r\geq\tilde r$ where $\tilde r$ is the solution
of the equation:
\bu
1-{2m\over\tilde r}+{Q_0^2\over\tilde r^2}=\left(1-{3m\over\tilde r}+{2Q_0^2
\over\tilde r^2}\right)^2
\lab{y47}
\eu
\begin{figure}[\ht]
\vspace{-0.4cm}
\centerline{\epsfig{figure=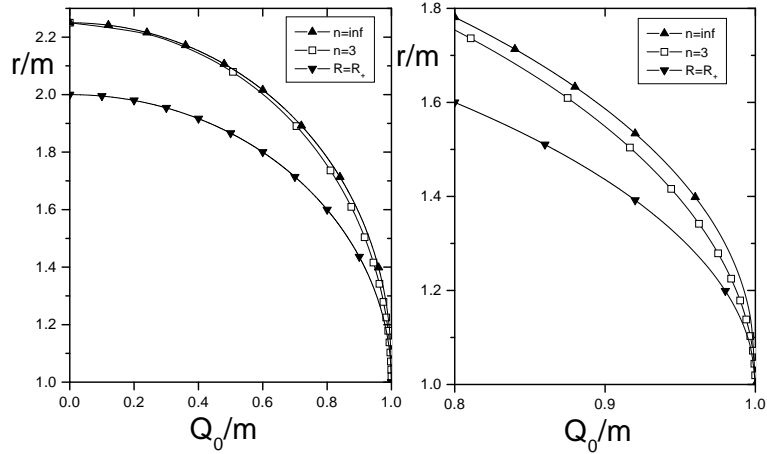, height=7cm}}
\caption{Configurations in the parameter space for different n and enlargement}
\label{Fig5}
\end{figure}

If $Q_0=0$, we easily deduce $\tilde r=9M/4$, a well known limit for regular 
internally static, electrically neutral, spherically symmetric and 
 perfect fluid solutions (Buchdahl, 1959). 
If $Q_0\not=0$, there are several internal solutions which match the 
external Reissner-Nordstr\"om one. Here, of the various solutions we 
shall consider those  with $n=3$ and $n=\infty$.
The plots of the limit of regularity for these solutions, namely where 
the central pressure diverges, are compared in the ($r-Q_0$)-plane and 
shown in
 Figure \re{Fig5} where $r$ is the coordinate radius of the sphere's boundary.
Although the two curves coincide at $r=9M/4$ when $Q_0=0$ 
and $r=M$ when $Q_0=M$,
as expected, they differ from each other everywhere else, 
 curve $n=\infty$ being at larger radii than 
curve $n=3$ for each value of $Q_0$. Now it happens that the condition 
for infinite central pressure in the $n=\infty$ case (a perfect conductor)
can be expressed analytically through solving the Oppenheimer-Wolkoff 
equation for conductive incompressible perfect fluids and reads:
\bn
\sqrt{1-{2m\over\tilde r}+{Q_0^2\over\tilde r^2}}&=&
{\tilde r m-Q^2_0\over 3\tilde r m-2Q^2_0}.
\lab{y48}
\en
This coincides with condition (\re{y47}) for regular embedding.
It is then clear that regular solutions for $n=3$ exist which 
have radii smaller than the corresponding limit of regular embedding; 
this is sufficient to falsify the conjecture. 
  
\section{Conclusions}

The use of plots in Figures \re{Fig1} to \re{Fig5} highlights the result
that regular solutions of Einstein's equations exist 
describing charged spheres arbitrarily close to a black-hole.
These solutions fill a region in a parameter space which is thin but 
finite so it is not a set of measure zero. 
However not all these solutions are stable against adiabatic radial 
perturbations. We have made  a perturbation analysis and have established a 
stability criterion given by relation (37). We have deduced the stability 
curves for configurations with adiabatic indeces $\gamma=5/3$ and 
$\gamma=4$ (respectively curves $e$ and $d$ in figures (1) and (2)), 
assuming that $\gamma$ was constant throughout the sphere.
We see that, for moderate values of $\gamma$, the limits of stability 
prevent stable compact charges from being arbitrarily close to a 
black-hole state 
but also from having a total charge $Q_0$ arbitrarily close to their mass. 
These situations would be possible if the adiabatic index diverged; in this 
case, in fact, the stability curve approaches curve $c$ which, we recall, is 
the limit of regularity for our solutions. Clearly, a diverging 
$\gamma$ corresponds to a complete rigidity while, in a less extreme case,
$\gamma$ would critically depend on the equation of state (Akmal et al., 1998).

Although our solutions can be thought of as being classic (non quantum) models 
of charged particles, it would be interesting to find solutions of Einstein's 
equations which describe charged spheres with a more detailed physical 
description of their interior.

\section{Acknowledgments}

 One of us (F.de F.) would like to thank the Director of the 
Physics Department of Peking University for hospitality.
We thank Prof. Mary Evans Properi for correcting the English text.

\end{document}